\documentclass[fleqn,twoside]{article}
\usepackage{espcrc2}

\usepackage{graphicx}
\usepackage[figuresright]{rotating}

\newcommand{\AmS}{{\protect\the\textfont2
  A\kern-.1667em\lower.5ex\hbox{M}\kern-.125emS}}

\title{String effects in Polyakov loop correlators.}

\author{M. Caselle\address[TO]{Dip. di Fisica Teorica, Universit\`a di
Torino and INFN, sezione di Torino,\\ 
Via P. Giuria 1, I-10125 Torino, Italy},
M. Hasenbusch\address[D]{NIC/DESY Zeuthen, Platanenallee 6, D-15738 Zeuthen,
 Germany},
M. Panero\addressmark[TO] and 
P. Provero\address{D.I.S.T.A., Universit\`a del Piemonte Orientale and
INFN,\\ 
gruppo collegato di Alessandria, I-15100 Alessandria,
Italy}\addressmark[TO]}

\begin{document}

\begin{abstract}
We compare the predictions of the effective string description of
confinement in finite temperature gauge theories to high precision
Monte Carlo data for the three-dimensional $Z_2$ gauge theory. 
We show that string interaction effects become more relevant as the
temperature is increased towards the deconfinement one, and are well
modeled by a Nambu-Goto string action.
\end{abstract}

\maketitle

\section{INTRODUCTION}
The effective string picture of confinement describes the quantum
fluctuations of the color flux tube that joins confined color sources
in gauge theories. By assuming that such fluctuations are described by
a bosonic string theory one obtains quantitative predictions about
gauge invariant correlation functions in the confined phase, that can
be compared to the results of Monte Carlo simulations of gauge
theories in the confining regime to obtain a stringent test of the
validity of such an effective description (see
Ref.\cite{Caselle:2002rm} for references to the original literature).

The three-dimensional $Z_2$ gauge theory is an ideal testing ground
for the effective string description of confinement: on one hand, 
its configuration
space is very small compared to the one of four dimensional Yang-Mills
theories, so that  high precision Monte Carlo results can be obtained 
with comparatively small computational effort; on the other hand it is
widely believed that the effective string theory that describes
confinement is to a large extent universal, that is independent from
the gauge group and the dimensionality of spacetime. Therefore many
results obtained in the context of $Z_2$ gauge theory are likely to be
valid with little or no modification in
Yang-Mills
theories as well.

In this contribution we compare the predictions of the effective
bosonic string theory of confinement to Monte Carlo results 
obtained in 
three dimensional $Z_2$ gauge theory at finite temperature. 
The gauge-invariant quantity
of interest is the correlation function of two Polyakov loops in the
confined phase. The effective string theory predicts the
dependence of this correlation function from the temperature, the
distance between the Polyakov loops, and the zero temperature string
tension. Our aim is to compare such predictions to Monte Carlo
results. First, however, a careful discussion of the physical regime
in which the string picture is expected to hold is in order: in
particular we will discuss what region in the parameter space is
expected to be described by a {\it free} bosonic string theory, and
where one should instead expect corrections to this picture. 

For a more detailed discussion of all these issues, see
Ref.\cite{Caselle:2002rm}. Note however that the numerical data reported in the
present contribution are much more precise than the ones reported in
Ref.\cite{Caselle:2002rm}, being obtained with a new algorithm for the
computation of Polyakov 
loop correlators that will be described in a forthcoming publication. 

\section{FREE STRING PREDICTIONS AND THEIR RANGE OF VALIDITY}

Suppose the quantum fluctuations of the color flux tube are modelled by
a free bosonic string theory. Then if we are interested in Polyakov
loop correlation functions, such a string lives on a world sheet that is
bounded by the two loops in the space-like direction, and periodic in
the time like direction. The partition function of such a string can
be computed, for example, using $\zeta$-function regularization. The
prediction for the Polyakov loop correlation function is 
\begin{equation}
\langle P(0)P^{\dagger}(R)\rangle=\exp\left[-F(R,L)\right]
\label{string1}
\end{equation}
where $L=1/T$ is the spatial size of the lattice, and $F(R,L)$, the
free energy of the string, is made of a classical and a quantum
contribution: in three spacetime dimensions
\begin{eqnarray}
F_{\rm cl}(R,L)&=&\sigma_0 L R +k(L)\label{string2}\\
F_{\rm q}(R,L)&=&\log\eta(\tau)  
\label{string3}
\end{eqnarray}
where $\sigma_0$ is the zero temperature string tension at the same
coupling, $k(L)$ is a non universal constant depending on $L$ only,
$\eta$ is the Dedekind function, and $\tau$ is the modular
parameter $\tau\equiv{iL}/{2R}$.

Physical considerations allow us to determine the range of physical
parameters where we expect the free string picture to hold: in
particular, both distances $R$ and $L$ appearing in Eq.\ref{string1}
must be large in the following sense: 
\begin{itemize}
\item 
{\it large} $R$: 
the picture of a thin, free string is certainly an idealization,
since we know that the actual flux tube has a non zero width of order
$1/T_c$ where $T_c$ is the deconfinement critical
temperature. Therefore we do not expect the prediction embodied by
Eqs.(\ref{string1}-\ref{string3}) to hold for distances $R<1/T_c$.
\item
{\it large} $L$: if one assumes the free string picture to be valid at
all temperatures up to $T_c$, one obtains a prediction for the value of
such temperature which is very far from the actual value obtained from
Monte Carlo simulations. The free string picture must break down for
temperatures close to $T_c$.   
\end{itemize}
\section{CORRECTIONS TO THE FREE STRING PICTURE: THE NAMBU-GOTO
STRING}
The free string described in the previous section is the infrared
limit of a large class of models having in common the bosonic
character of the field describing the fluctuations of the color flux
tube (see \cite{Dietz:1983uc}). 
Among these various models, the Nambu-Goto string
is the best candidate to describe the short distance corrections to
the free string picture that, as discussed above, must be present in
the effective description of confinement. 

There are at least two
reasons for choosing this particular model: first, it gives a
prediction for the dimensionless ratio $T_c/\sqrt{\sigma_0}$ which is
in good agreement with Monte Carlo results for several gauge
models. Second, the Nambu-Goto string accurately describes interface
fluctuations in the three-dimensional Ising model
\cite{Caselle:1994df}, 
which is related by
duality to the $Z_2$ gauge theory. For these reasons, we will use the
Nambu-Goto string to model short distance corrections to the
free string picture.

For Polyakov loop correlators, the Nambu-Goto string predicts a
correction to the free string behavior that has been computed in
\cite{Dietz:1983uc} using $\zeta$-function regularization at two-loop level, 
(the
loop expansion parameter being $1/(\sigma_0 L R)$) and is given, in three
dimensions,  by:
\begin{equation}
F_q^{(NLO)}=-\frac{\pi^2 L}{1152\ \sigma_0
R^3}\left[2
E_4(\tau)-E_2^2(\tau)\right]
\label{nlo}
\end{equation}
in terms of the Eisenstein functions $E_2$ and $E_4$. Note that these
corrections do not involve any free parameters.

\section{RESULTS}
The comparison between the string predictions described above and
Monte Carlo data for Polyakov loop correlation functions is shown in
Figs. 1 and 2. They refer to the same value of the coupling
$\beta=0.75180$, corresponding to a critical temperature $T_c=1/8 a$
\cite{Caselle:1995wn}, where 
$a$ is the lattice spacing. The size $L$ of the lattice for the two
figures is 24 and 16, so that the temperatures are
respectively $T_c/3$ and $T_c/2$. The Monte Carlo data plotted
correspond to the quantity
\begin{equation}
Q_q(R,L)=\log\left(\frac{G(R)}{G(R+1)}\right)-\sigma_0 L
\end{equation}
as a function of $z=2R/L$;
the zero-temperature string tension $\sigma_0$ is taken from
previously published high precision measurements
\cite{Hasenbusch:1997} taken 
in the dual Ising model. 

This quantity is compared to the free string prediction and to the
Nambu-Goto prediction. The figures show that:
\begin{itemize}
\item
As predicted, the free string picture describes the data better at
$T=T_c/3$ than at $T=T_c/2$
\item
With the high precision of our data, even at temperatures as low as
$T_c/3$ there are sizable corrections to the free string picture.
\item
Such corrections are very well modeled, both at $T_c/3$ and $T_c/2$, 
by the Nambu-Goto string action at
next-to-leading order
\end{itemize}
\begin{figure}
\includegraphics[width=7.5cm]{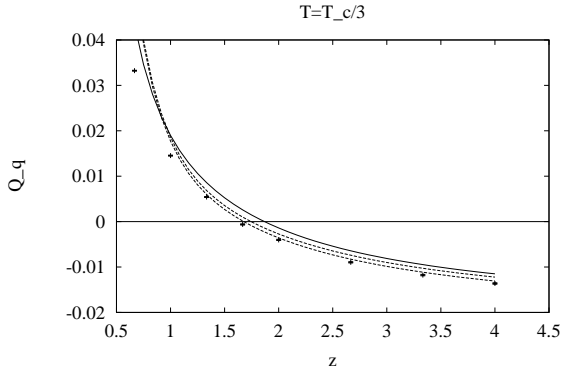}
\vskip-0.5cm\caption{\sl Data at $T=T_c/3$: the solid line is the free string
prediction; the dotted lines enclose the range of predictions given by
the Nambu Goto string (the uncertainty derives from the uncertainty on
$\sigma_0$).}
\end{figure}
\begin{figure}
\includegraphics[width=7.5cm]{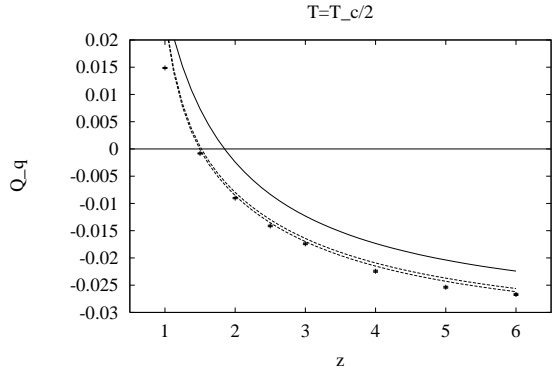}
\vskip-0.5cm\caption{\sl Same as Fig. 1 for $T=T_c/2$} 
\end{figure}
\section{CONCLUSIONS}
Confinement in the three dimensional $Z_2$ gauge theory is mediated by
a flux tube which fluctuates like a bosonic string. This fact was
established for the zero temperature case by studying finite size
effects in Wilson loop expectation values \cite{Caselle:1996ii}: the results
presented here show that this is the case also at finite
temperature. Moreover, we have established that corrections to the
free string behavior are present, as expected, at high temperatures,
and are well modeled by a Nambu-Goto string action.

\end{document}